\documentclass[12pt]{article}
\usepackage{amssymb,amsmath,amscd,latexsym,array}
\newcommand{\be}{\begin{equation}}
\newcommand{\ee}{\end{equation}}
\newcommand{\bea}{\begin{eqnarray}}
\newcommand{\eea}{\end{eqnarray}}

\newcommand{\ba}{\begin{array}}
\newcommand{\ea}{\end{array}}

\newcommand{\bt}{\begin{tabular}}
\newcommand{\et}{\end{tabular}}

\newcommand{\fr}{\frac}
\newcommand{\ci}{\cite}
\newcommand{\cl}{\centerline}
\newcommand{\bs}{\bigskip}
\newcommand{\vs}{\vspace}
\newcommand{\hs}{\hspace}

\newcommand{\en}{\eqno}

\newcommand{\fnt}{\footnotetext}
\newcommand{\fm}{\footnotemark}
\newcommand{\bbib}{\begin{thebibliography}}
\newcommand{\ebib}{\end{thebibliography}}

\newcommand{\bi}{\bibitem}

\begin{document}

\hs{4.5cm}{\it LANDAU INSTITUTE PREPRINT 23/11/98}

\vs{2cm}
\centerline{\bf NIEMEIER SELF-DUAL LATTICES}
\cl{\bf AND TOPOLOGICAL  PHASE TRANSITIONS}

\bigskip

\centerline{\bf  S.A.Bulgadaev\fm{}}\fnt{This work is supported
by RFBR grants 96-02-17331 and 96-1596861}

\bigskip

\centerline{ Landau ITP RAS, 117334, Moscow,  Kosyghin street, 2}

\cl{E-mail address: bulgad@itp.ac.ru}

\vs{1cm}

\cl{Abstract}

\vspace {1.0cm}

A topological phase transition in two-dimensional
nonlinear $\sigma$-models on tori, connected
with self-dual (unimodular) 24-dimensional Niemeier lattices,
is considered.
It is shown  that
critical properties of these transitions are determined by
corresponding Coxeter numbers of lattices. A case of general
integer-valued lattices with minimal norm equal 1 and 2 and
a possible application to string theory are discussed.

\newpage

Recently it was shown \ci{1} (see also \ci{2}), that
two-dimensional nonlinear
$\sigma$-models, defined on the maximal abelian Cartan tori $T_G$
of simple compact  Lie groups $G$, have a topological phase transition
with critical properties different from that of $XY$-model or
of nonlinear $\sigma$-model on circle $S^1$ \ci{3,4}
There are the universality  classes of critical behaviour,
determined by the type of the corresponding
dual root lattices $L_v$, which can belong to the series
$A_n, D_n, E_n, Z_n.$  More rigorously, all critical indices
are determined by a set of minimal vectors and
can be expressed through
the Coxeter number $h_G$ of these lattices.
For example, a correlation lentgh $\xi$ as a function of temperature T
(or equivalent coupling constant of the nonlinear $\sigma$-model)
has an essential singularity
$$
\xi(T) \sim a \exp (A\tau^{-\nu_G}), \, \tau = \fr{T-T_c}{T_c} ,
\en(1)
$$
$$
\nu_G = 2/(h_G +2),
\en(2)
$$
where $a$ is a small size cut-off parameter, $A$ is some nonuniversal constant
$O(1)$ and $T_c$ is a
temperature of phase transition. To the initial  $XY$-model, usual tori
$T^n=U(1)^n$ and $Z^n$ lattices corresponds
$h_G=h_{A_1}=2.$

The Coxeter numbers of $A,D,E$ root systems have the following values
$$
h_{A_n}= n+1, \, h_{D_n} = 2n-2, \, h_{E_6} = 12,\, h_{E_7} =18, \,
h_{E_8}= 30
\en(3)
$$

It means that in this relation  tori $T_G,\, G=A,D,E,$ cannot be considered
as a direct product of $S^1 = U(1)$: $T_G \ne U(1)^n.$

All tori $T_G$ are determined by so called integer-valued lattices
(in appropriate scale) of $A,D,E,Z$ series.
A series $Z_n$  belongs to the odd self-dual (or unimodular) lattices
and contains a minimal vectors with norm equal 1, while the series
$A_n, D_n, E_n$ belong to the even lattices with minimal norm equal 2.
Among them only lattice $E_8$ is self-dual. There are a number of
other integer-valued  lattices, but the most interesting are self-dual
ones. Besides
possible application of these lattices in a construction of the
anomaly-free string theory \ci{5} one can show that
for such lattices and connected with them tori and nonlinear $\sigma$-models
all quantum numbers admit a topological interpretation: they can be
represented as a topological charges of classical solutions \ci{6}.

From the point of view of topological phase transitions the most
importance have  the integer-valued lattices with minimal norm equal 2,
since, in principle, only they can have new  critical properties.
But their number and classification are not known at present (except some
low-dimensional cases). If for this reason we confine ourselves only
by self-dual
lattices, even then their number is too large and grows rapidly with
lattice dimension $d$. For example, a number of such lattices in dimensions
$d\le 23$ is 106 \ci{8}. Thus we shall consider only even self-dual lattices.
They can exist only in
spaces with dimension $d= 8n,$ $n\in Z^{+}$ \ci{7}. In $d=8$ there is
one such lattice $E_8,$ in $d=16$ there are two lattices:
$E_8 \oplus E_8$ and $D_{16}^{+}.$ Here a lattice $D_{16}^{+}$ is obtained
from lattice $D_{16}$ by addition one gluing vector \ci{8}, but its
set of the minimal vectors coincides with that of lattice $D_{16},$
and for this reason their critical properties also coincides.

In space with $d=24$ there are 24 self-dual even lattices, enumerated
and constructed by
Niemeier \ci{9} and 23 from them have minimal norm equal 2 (note that
a number of such odd lattices is much larger and equals 155 \ci{8}).
The last lattice,
the so called Leach lattice, has a minimal norm equal 4 \ci{10}. In the next
space with $d = 32$ there are no less than $8\cdot 10^7$ even unimodular
lattices \ci{8} and for this reason any analysis of such lattices
in spaces with $d>24$ is now also impossible.

In this paper
we consider critical properties of nonlinear $\sigma$-models, connected
with 23 even unimodular lattices with a minimal norm equal 2,
which we will call as a $N_n$ $(n=1,...,23)$ series.
All lattices of this series
are constructed by a gluing method from lattices of $A,D,E$ series,
such that all components must have the same Coxeter number $h_n$ and a sum
of their dimensions must be equal to 24 \ci{8,9}.
In analogy with lattices of $A,D,E$ type a set of minimal vectors with
norm equal 2 are named the roots of these lattices.

A geometry of the root systems $\{r\}_n$ of the Niemeier lattices,
obtained by using some theorems
about dimension of the modular forms space, is the following \ci{8}:

1)a rank (or dimension $d$) of the root set is equal 24;

2)a number of roots in $N_n$ lattices are equal to $24 h_n$;

3)an isotropy constant $B_n$ of the root systems $\{r\}_n$, defined by condition
$$
B_n \delta_{ik} = \sum_{\{r^a\}} r_i^a r_k^a,
\en(4)
$$
is equal
$$
B_n=2h_n
\en(5)
$$

Compare this with corresponding data for root systems of $A,D,E$
lattices one can show that the root systems of $N_n$ lattices
break into direct sum of the root systems of the lattices,
composing a lattice $N_n,$ and that the Coxeter numbers
for $N_n$ lattices has the same property as the Coxeter numbers
of the $A,D,E$ lattices, i.e a number of all roots $\#_n$ in $N_n$ lattice
is equal $\#_n = n h_n.$
Now one can enumerate all Niemeier lattices, their number is
a number of partitions of $d=24$ into the dimensions of the root systems
of $A,D,E$ types
with condition, that all components must have the same Coxeter number.

Taking into account (3) one can show that there are only 23 combinations.
They are listed together with their Coxeter numbers in a Table 1 \ci{8}

\bs

\cl{Table 1}
\vs{0.4cm}
{\small
$$
\ba {|c|c|c|c|c|c|c|}
\hline
N_n & 24A_1 & 12A_2 & 8A_3 & 6A_4 & 6D_4 & 4A_5D_4\cr
\hline
h_{N_n}& 2&3&4&5&6&6\cr
\hline
\ea
$$
$$
\ba {|c|c|c|c|c|c|c|}
\hline
N_n & 4A_6 & 2A_72D_5 & 3A_8 & 4D_6 & 2A_9D_6 & 4E_6 \cr
\hline
h_{N_n}&7&8&9&10&10&12\cr
\hline
\ea
$$
$$
\ba {|c|c|c|c|c|c|}
\hline
N_n & A_{11}D_7E_6 & 2A_{12} & 3D_8 & A_{15}D_9 & D_{10}2E_7 \cr
\hline
h_{N_n}& 12&13&14&16&18\cr
\hline
\ea
$$
$$
\ba{|c|c|c|c|c|c|c|}
\hline
N_n & A_{17}D_7 & 2D_{12} & A_{24}&3E_8 & D_{16}E_8&D_{24}\cr
\hline
h_{N_n} &18&22&25&30&30&46\\
\hline
\ea
$$}

\vs{0.4cm}
\noindent
Here a symbol $iA_rkD_slE_t$ denotes that the corresponding $N_n$ lattice
is composed
from i$A_r$, k $D_s$ and l $E_t$ lattices.
As was shown in \ci{1} all coefficients of renorm-group equations,
determining critical properties of the models, are expressed through
Coxeter number. Since all components of each $N_n$ lattice have
the same Coxeter numbers it means that the nonlinear
$\sigma$-models, related with the
Niemeier lattices,  will have the same critical properties as
the $\sigma$-models related with $A,D,E$ lattices with the same
Coxeter numbers!
Note also, that again, as in the case of $A,D,E$ lattices,
there are
some degenerations between different  $N_n$ lattices on the Coxeter
number (see Table 1) and consequently all critical properties
of the corresponding models will coincide.

Thus we have shown that no any new universality classes appear in the case
of 24-dimensional even self-dual lattices with minimal norm equal 2.
It turns out that this fact takes place for all integer-valued
lattices with minimal norm equal 1 or 2. This conclusion follows from
the Witt theorem \ci{8}, proving that
the set of their minimal vectors must be a direct sum of the root systems.
But they can be only of $A,D,E,Z$ types. Consequently, all integer-valued
lattices with minimal norm equal 1 can have critical properties only of
$XY$-model (or of $Z^n$ lattice) type, while all integer-valued lattices
with minimal norm equal 2 can have critical properties only of $A,D,E$
lattice types. Now different components can have different Coxeter
numbers and one can have a series of phase transitions in each component
(discussion of some simple examples see below).

Due to existence of low-temperature massless phase a topological
phase transition in nonlinear $\sigma$ models on tori can be
interpreted as a {\it statistical decompactification} transition \ci{11}.
It means that this transition does not really change a topology of the target
space $T^d$, but only effectively, i.e. it changes a behaviour of all
correlations at large distancies
to those in free theory defined on covering target space $R^d$.
From this point of view such topological phase transition can have
some interest for string theories.
As we have shown a phase transition in $\sigma$-models on composite
torus takes place simultaneously
in the whole torus space,
if all components have the same Coxeter number.
For string theory a more interest would have a possibility of topological
phase transition of above mentioned type from some compact space
${\cal C}$ into partially
decompactified space of the form
$$
{\cal M}= R^{d}\otimes {\cal C}',
$$
where ${\cal C}'$ is another compact space, which can be identified with
internal (isotopic) space. For this one need to consider a $\sigma$-model
on a target space ${\cal C}$ with homotopical group $\pi_1=L^d,$ where
$L^d$ is some $d$-dimensional lattice of $A,D,E,Z$ type. In general,
${\cal C}$ can contain some torus ${\cal T}^d$ as a submanifold and
has corresponding relative homotopical group \ci{6,11}
$$
\pi_2({\cal C},{\cal T}^d)=L^d.
$$
Below, for simplicity,
we suppose that
$${\cal C} = {\cal T}^d \otimes {\cal C}'.
$$
Here ${\cal T}^d$ is some $d$-dimensional torus, corresponding
to the lattice $L^d.$ Let us first consider a case
$\pi_1({\cal C}')=0.$
Then topological phase transition will effectively
decompactified torus and target space becomes ${\cal M}.$
Another possibility is when ${\cal C}$ is a composite torus
$${\cal T} = {\cal T}_1 \otimes {\cal T}_2.$$
Then one can have two (or $n$, if a number of components is equal $n$)
phase transitions at  different temperatures
$T_{ci}$ in $i$-th component of the composite torus,
under this a  compact composite torus ${\cal T}$
will behave itself at intermediate temperatures effectively as
a partially noncompact space
$$
{\cal M}' = R^{d_2} \otimes {\cal T}_1,
$$
where $R^{d_2}$ corresponds
to the decompactified component ${\cal T}_{2}.$
This scenario is possible when components of the composite torus ${\cal T}$
have different Coxeter numbers $h_i$ and, for example,
$h_1 > h_2.$

The corresponding schematic phase diagram
is depicted in Fig.1. Here $g$ is a dimensionless amplitude
of one vortex, $\delta$ is dimensionless parameter, connected with coupling
constant and $\tau.$  Lines 1,2 denote lines of phase transitions
in torus ${\cal T}_1$ and torus ${\cal T}_2$ respectively,
$X$ denotes an intermediate phase with partial decompactification:
torus ${\cal T}_2$  is decompactified, torus ${\cal T}_1$ still not.
Line 3 denotes a symmetric separatrix,  which lies in massive phase,
its incline is universal for
all tori and on it there is an additional symmetry. For tori $T_G$ it is
a symmetry of the corresponding
continuous compact Lie group $G$ \ci{12}.

\begin{picture}(400,150)(-75,0)
\put(100,0){\vector(1,0){100}}
\put(100,0){\vector(0,1){100}}
\put(100,0){\line(-1,0){100}}
\put(100,0){\line(1,1){80}}
\put(100,0){\line(-1,1){80}}
\put(100,0){\line(2,1){80}}
\put(170,20){low T phase}
\put(58,60){massive phase}
\put(210,0){$\delta$}
\put(100,110){g}
\put(185,37){1}
\put(10,80){3}
\put(170,50){X}
\put(185,80){2}
\put(100,-15){0}
\end{picture}
\vspace{0.5cm}

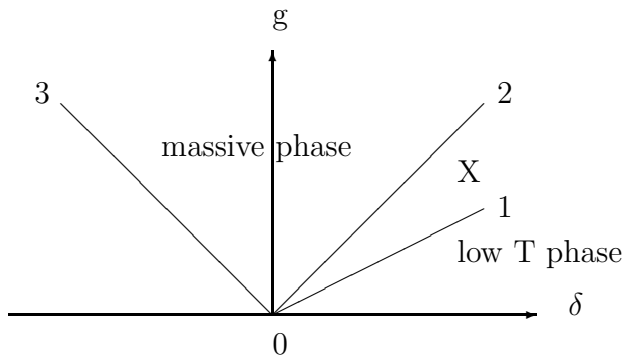
\cl{ Fig.1. Schematic phase diagram of $\sigma$-model}

\cl{on composite torus ${\cal T}= {\cal T}_1 \otimes {\cal T}_2$}

\vs{0.5cm}

It follows from (3) that the Coxeter numbers $h_G$ increase
with increase of dimension of lattices and for lattices with equal
dimensions the next inequalities take place
$$
h_{A_n} \le h_{D_n} < h_{E_n}.
\en(6)
$$
Thus, if both tori belong to the same series, firstly a torus with smaller
dimension must be decompactified.  When one torus
is of usual type ${\cal T}_2=R^{d_2}/L^d$ ($h=2$ always corresponds to them)
and other torus is of $T_G \, (G=A,D,E,)$ type with $d_1>1$, then the usual
torus will decompactify first. For obtaining 4-dimensional space $R^4$
(we confine here ourselves only by Euclidean case) one need take a torus
$T_G$ of one of the simple compact Lie groups with rank equal 4.
Among them there is  $SU(5)$ group, which is used often as a group of
Grand Unification Theories! Moreover, as was noted in \ci{11}, if we
consider analogous decompactification transition in chiral model on
$G=SU(5),$ we can obtain as an internal space a corresponding flag space
$F_G=G/T_G$ with dimension $d=20$ and  with homotopical group
$\pi_2(F_G)=L_v^G$ ($L_v^G$ is a lattice of dual roots of group $G$),
 which is needed
for topological interpretation of the quantum numbers of group $G$ \ci{6}.
It is worth to note that quantum numbers of the fundamental (quark)
representations
of groups $SU(N)$ do not admit topological interpretation in terms
of instantons on $F_G$ and maybe for this reason the quarks do not freely
exist.

Spaces ${\cal M}'$ are also analogous to the partially compactified
spaces of the string theory, but here they appear as an {\it intermediate}
spaces. Under further cooling, in low $T$ phase, both torus become effectively
decompactified.
A physical picture of decompactification transition in chiral models on tori
is very attractive. It can serve as a guide line for more realistic
string models, which must {\it describe a process of formation of the
partially compactified space}. In our opinion these processes must be
similar to the cosmological processes of the "birth of Universe" type
and for this reason must take into account the cosmological aspects as well
as the fact that string partially compactified
spaces are the final (or even maybe the intermediate) spaces of the
general process of the Universe development.

\bbib {50}
\bi{1} Bulgadaev S.A., Pisma v ZETP, {\bf 63} (1996) 743,
			 Landau Institute Preprint, 1997.
\bi{2} Bulgadaev S.A., Phys.Lett., {\bf 86A} (1981) 213.
\bi{3} Berezinsky V.L., ZETP, {\bf 59} (1970) 907, ibid. {\bf 61} (1971) 114.
\bi{4} Kosterlitz J.M., Thouless D.J., J.Phys., {\bf C6} (1973)118;
			 Kosterlitz J.M., J.Phys., {\bf C7} (1974) 1046.
\bi{5} Green M.B., J.H.Schwarz, E.Witten. Superstring theory, v.1,2,
Cambridge University Press, 1988
\bi{6} Bulgadaev S.A., Pisma v ZETP, {\bf 63} (1996) 758.
\bi{7} Hecke E., Math.Ann., {\bf 114} (1937) 1-28.
\bi{8} Conway G,H., Sloane N.J.A., Sphere Packings, Lattices and Groups,
v.I,II. Springer-Verlag, 1988.
\bi{9} Niemeier H.-V., JNT, {\bf 5} (1973) 142 -178.
\bi{10} Leech J., Can.J.Math., {\bf 19} (1967) 251-267.
\bi{11} Bulgadaev S.A., On decompactification transition in
two-dimensional chiral models. Talk given at International conference
"Conformal field theory and Integrable models", June 1996, Chernogolovka;
 Landau Institute Preprint, 1997.
\bi{12} Bulgadaev S.A., Nucl.Phys., {\bf B224} (1983) 349;
Topological transitions in two-dimensional systems with internal symmetries.
Landau Institute preprint, 1997.

\ebib

\end{document}